\documentclass[aps,prl,twoside,twocolumn,floatfix,superscriptaddress,amsmath,showpacs,amssymb]{revtex4}
\usepackage{amssymb}
\usepackage{bm}
\usepackage{graphicx}
\usepackage{psfrag}
\usepackage{units}
\usepackage{lineno}
\usepackage{color}

\def\buch{Institute for Nuclear Physics and Engineering, Bucharest, Romania}
\def\buda{KFKI Research Institute for Particle and Nuclear Physics, Budapest,
  Hungary}
\def\cler{Laboratoire de Physique Corpusculaire, IN2P3/CNRS, and
  Universit\'{e} Blaise Pascal, Clermont-Ferrand, France}
\def\sp{University of Split, Split, Croatia}
\def\darm{Gesellschaft f\"{u}r Schwerionenforschung, Darmstadt, Germany}
\def\dres{Institut f\"{u}r Strahlenphysik, Forschungszentrum
  Dresden-Rossendorf, Dresden, Germany}
\def\heid{Physikalisches Institut der Universit\"{a}t Heidelberg, Heidelberg,
  Germany}
\def\mosc{Institute for Theoretical and Experimental Physics, Moscow, Russia}
\def\kurc{Kurchatov Institute, Moscow, Russia}
\def\seou{Korea University, Seoul, Korea}
\def\stra{Institut Pluridisciplinaire Hubert Curien and Universit\'{e} Louis
  Pasteur, Strasbourg, France}
\def\wars{Institute of Experimental Physics, University of Warsaw, Warsaw, Poland}
\def\zagr{Ru{d\llap{\raise 1.22ex\hbox
  {\vrule height 0.09ex width 0.315em}\kern 0.04em}}er
  Bo\v{s}kovi\'{c} Institute, Zagreb, Croatia}

\def\lan{Institute of Modern Physics, Chinese Academy of Sciences, Lanzhou,
  China}

\def\mun{Physik Department, Technische Universit\"{a}t M\"{u}nchen, Germany}
\def\jap{Heavy-Ion Nuclear Physics Laboratory, RIKEN, Wako, Japan}
\def\vien{Institut f\"{u}r Mittelenergie-Physik, \"{O}sterreichische Akademie
der Wissenschaften, Wien, Austria}
\def\frank{Frankfurt Institute for Advanced Studies,
           Johann Wolfgang Goethe University, Frankfurt am Main, Germany}
\def\giess{Institut f\"{u}r Theoretische Physik, Giessen, Germany}

\begin{document}

\title{Measurement of the in-medium $K^0$ inclusive cross section
   \\ in $\pi^{-}$-induced reactions at 1.15~GeV/c}

\author{M.L.~Benabderrahmane} \affiliation{\heid}
\author{N.~Herrmann} \email{herrmann@physi.uni-heidelberg.de} \affiliation{\heid} 
\author{K.~Wi\'{s}niewski} \affiliation{\wars}
\author{J.~Kecskemeti} \affiliation{\buda}
\author{A.~Andronic} \affiliation{\darm}
\author{V.~Barret} \affiliation{\cler}
\author{Z.~Basrak} \affiliation{\zagr}
\author{N.~Bastid} \affiliation{\cler}
\author{P.~Buehler} \affiliation{\vien}
\author{M.~Cargnelli} \affiliation{\vien}
\author{R.~\v{C}aplar} \affiliation{\zagr}
\author{E.~Cordier} \affiliation{\heid}
\author{I.~Deppner} \affiliation{\heid}
\author{P.~Crochet} \affiliation{\cler}
\author{P.~Dupieux} \affiliation{\cler}
\author{M.~D\v{z}elalija} \affiliation{\sp}
\author{L.~Fabbietti} \affiliation{\mun}
\author{Z.~Fodor} \affiliation{\buda}
\author{P.~Gasik} \affiliation{\wars}
\author{I.~Ga\v{s}pari\'c} \affiliation{\zagr}
\author{Y.~Grishkin} \affiliation{\mosc}
\author{O.N.~Hartmann} \affiliation{\mun}
\author{K.D.~Hildenbrand} \affiliation{\darm}
\author{B.~Hong} \affiliation{\seou}
\author{T.I.~Kang} \affiliation{\seou}
\author{P.~Kienle }  \affiliation{\vien} \affiliation{\mun}
\author{M.~Kirejczyk} \affiliation{\wars}
\author{Y.J.~Kim} \affiliation{\darm}
\author{M.~Ki\v{s}} \affiliation{\darm} \affiliation{\zagr}
\author{P.~Koczo\'n} \affiliation{\darm}
\author{M.~Korolija} \affiliation{\zagr}
\author{R.~Kotte} \affiliation{\dres}
\author{A.~Lebedev} \affiliation{\mosc}
\author{Y.~Leifels} \affiliation{\darm}
\author{X.~Lopez} \affiliation{\cler}
\author{V.~Manko} \affiliation{\kurc}
\author{J.~Marton} \affiliation{\vien}
\author{A.~Mangiarotti} \affiliation{\heid}
\author{M.~Merschmeyer} \affiliation{\heid}
\author{T.~Matulewicz} \affiliation{\wars}
\author{M.~Petrovici} \affiliation{\buch}
\author{K.~Piasecki} \affiliation{\heid} \affiliation{\wars}
\author{F.~Rami} \affiliation{\stra}
\author{A.~Reischl} \affiliation{\heid}
\author{W.~Reisdorf} \affiliation{\darm}
\author{M.~Rogowska} \affiliation{\wars}
\author{M.S.~Ryu} \affiliation{\seou}
\author{P.~Schmidt} \affiliation{\vien}
\author{A.~Sch\"{u}ttauf} \affiliation{\darm}
\author{Z.~Seres} \affiliation{\buda}
\author{B.~Sikora} \affiliation{\wars}
\author{K.S.~Sim} \affiliation{\seou}
\author{V.~Simion} \affiliation{\buch}
\author{K.~Siwek-Wilczy\'{n}ska} \affiliation{\wars}
\author{V.~Smolyankin} \affiliation{\mosc}
\author{K.~Suzuki} \affiliation{\mun}
\author{Z.~Tymi\'{n}ski} \affiliation{\wars}
\author{E.~Widmann} \affiliation{\vien}
\author{Z.G.~Xiao} \affiliation{\heid}
\author{T.~Yamazaki} \affiliation{\jap}
\author{I.~Yushmanov} \affiliation{\kurc}
\author{X.Y.~Zhang} \affiliation{\lan}
\author{A.~Zhilin} \affiliation{\mosc}
\author{J.~Zmeskal} \affiliation{\vien}
\collaboration{FOPI Collaboration} \noaffiliation
\author{E.~Bratkovskaya }  \affiliation{\frank}
\author{W.~Cassing} \affiliation{\giess}

\begin{abstract}

The $K^0$ meson production by $\pi^-$ mesons of 1.15~GeV/c
momentum on {\color{black} C, Al, Cu, Sn and Pb} nuclear targets was measured with the FOPI
spectrometer at the SIS accelerator of GSI. Inclusive production
cross-sections and the momentum distributions of $K^0$ mesons are
compared to {\color{black} scaled elementary production cross-sections and to 
predictions of theoretical models describing the in-medium production of kaons.
The data represent a new reference for those models, which are widely used 
for interpretation of the strangeness-production in heavy-ion collisions. 
The presented results demonstrate the sensitivity of the kaon production 
to the reaction amplitudes inside nuclei} and point to the
existence of a repulsive KN\,-\,potential of 20$\pm$ 5 MeV at normal
nuclear matter density.
\end{abstract}


\maketitle


Modifications of hadron properties in dense baryonic matter are a
subject of intensive research in hadron
physics~\cite{BROWN02}. Various theoretical
approaches~\cite{theory} agree qualitatively on predicting, for
example, modifications of masses and coupling constants for kaons
and anti-kaons. Due to the density dependence of the
$K$N($\overline{K}$N) potential, the $K^-$ effective mass is
expected to drop, whereas the mass of $K^+$ mesons is predicted to
rise with increasing density of nuclear matter. 
Due to additional attractive interactions   with the surrounding
nucleons a condensation of anti-kaons ($K^-$) may take place in a
dense baryonic environment as encountered in the interior of
neutron stars~\cite{Kaplan86}. 
Kaons ($K^+$, $K^0$), on the other hand, have a relatively long mean
free path in the nuclear matter at low momenta~\cite{Dover82}.
Therefore they are a good probe for studying the in-medium
properties of hadrons produced in collisions between nuclei at
energies close to the respective nucleon-nucleon production
thresholds~\cite{Fuchs06}. For an understanding of the
strangeness production in such collisions the knowledge of the
elementary production cross-sections at finite baryonic densities
is essential.

As far as the production of strangeness by pions is concerned,
there exist interesting predictions based on Quark-Meson Coupling (QMC)
model calculations~\cite{Tsushima00}, in which kaons ($K^+$,
$K^0$) and hyperons ($\Lambda$, $\Sigma$) are produced via the
formation of intermediate $\Delta$ and $N^*$ resonances. Due to
the in-medium modifications of the involved resonances, the
reaction amplitudes of these processes are expected to be
modified, 
{\color{black} thus giving rise to substantial changes of the kaon 
production cross sections at normal nuclear matter density, $\rho_0$.}

In this letter we report about measurements of the $K^0$ production 
by pions of 1.15 GeV/c momenta on various nuclear targets. The
inclusive production cross-sections and the $K^0$ phase-space
distributions are compared to theoretical predictions.
Evidence for the existence of 
a repulsive, mean-field KN\,-\,potential is presented.


The experiments were performed with the FOPI spectrometer at the
pion beam-line of the SIS accelerator at GSI. 
The  $\pi^-$ - beam had an intensity of about
3000~$\pi^-$~/s, a mean momentum of
1.15~GeV/c with a momentum dispersion of about 0.5~\%. 
The chosen beam momentum corresponds to an available energy $\sqrt{s}$
of about 1.75~GeV in the system of $\pi^-$ mesons colliding with 
nucleons at rest. The identification of charged particles is
achieved in FOPI~\cite{Gobbi92} by the curvature of particle
tracks in the magnetic field, by their specific energy loss in the
drift chamber and by the time of flight. In the present analysis
the geometrical acceptance was restricted to the polar angles
$25^{\circ}$~$<$~$\theta$~$<$$150^{\circ}$ covered by the
central drift chamber (CDC). $K^0_S$ mesons (c$\tau$~=~2.68~cm)
were reconstructed via their decays into ($\pi^-$, $\pi^+$) pairs,
whereas $K^0_L$ mesons could not be reconstructed due to their 
long life time. Five different
targets were used: C, Al, Cu, Sn, and Pb, with thicknesses of
1.87, 1.56, 4.41, 2.83 and 5.76~${\text{g/cm}}^{2}$, respectively.
Altogether about 25 million events were registered under a
minimum-bias trigger condition, i.e.\ in case that at least a
single charged particle was detected inside the CDC. The position
of the primary interaction point was reconstructed with the help
of two silicon micro-strip detector stations, each consisting of 
two single sided detectors
(3.2$\times$3.2${\text{cm}}^2$ area, 300~$\mu{\text{m}}$ in
thickness and with 50~$\mu$m pitch size) 
and placed 94~cm and 224~cm upstream from the target.

\begin{figure}[!th]
\vspace{-0.3cm}\includegraphics[width=7.6cm]{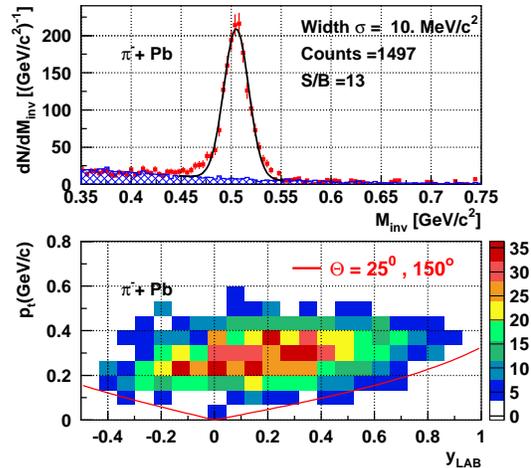}
\caption{\label{fig:K0_minv}Upper plot: the $K^0_S$ invariant mass
  distribution in $\pi^-+{\text{Pb}}$ reactions. 
{\color{black} Lower plot: Geometrical acceptance of the detector in the $\pi^-+{\text{Pb}}$
experiment in terms of rapidity versus transverse momentum of reconstructed 
$K^0_S$\,-\,mesons.}
 }
\end{figure}

The upper plot in Fig.~\ref{fig:K0_minv} shows the invariant mass distribution of
$(\pi^-,\pi^+)$ pairs registered in $\pi^-+{\text{Pb}}$ reactions.
The combinatorial background below the $K^0_S$ peak was
reduced by imposing selection criteria on parameters of the
reconstructed tracks: transverse momenta
($p_t(\pi^-,\pi^+)~\geq$~80~MeV/c), assigned masses of particles
(0.05~$\leq$~m($\pi^-,\pi^+$)~$\leq$~0.6~GeV/${\text{c}}^2$),
distances of closest approach to the primary vertex
($|d_0(\pi^-,\pi^+)|~\geq 1.5$~cm) and 
{\color{black} the differences between the azimuthal angles of reconstructed 
secondary vertices and kaon momenta} 
($|\Delta\phi|<30^\circ$). 
After
subtraction of the background - reconstructed by the event mixing
method - the distribution is fitted with a Gaussian function. A total
number of about 1500 $K^0_S$ is identified in the interval of \mbox{$\pm ~ 
2\sigma$} around the $K^0_S$ nominal mass, where the ratio of the
signal to the background is about 13.
Similar statistics is available for the C target, while for each of the 
other  targets  about 300 $K^0_S$ mesons are reconstructed.

\begin{figure}[!th]
\vspace{-0.3cm}\includegraphics[width=7.6cm]{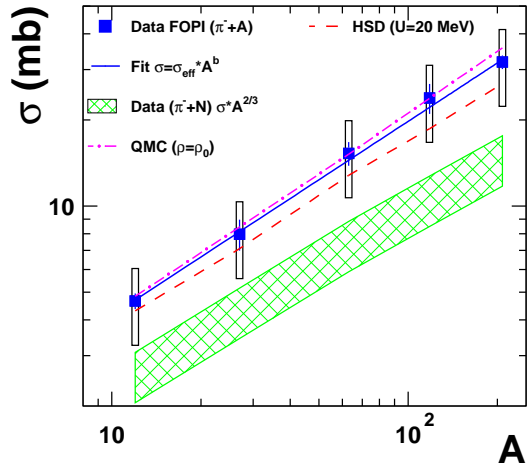}
\caption{\label{fig:cros_comp}The $K^0$ inclusive production cross
  section (squares) as a function of the mass number of the target nucleus. The solid 
  line represents  the fit with a power law function.
  The hatched area corresponds to the sum of the cross-sections of the
  elementary processes scaled according to the transverse size of the target
  nuclei.
  QMC model predictions at $\rho =\rho_0$~\cite{Tsushima00} (dashed dotted line) 
  are scaled with the same prescription, 
  whereas HSD transport model calculations (dashed line) yield absolute predictions.}
\end{figure}

Taking into account the geometrical acceptance of the apparatus
{\color{black} (shown in the lower plot in Fig. 1)}, 
the reconstruction efficiency of $K^0_S$ mesons,
the normalization to the number of beam particles, the target thickness
and the branching ratio into  $K^0_{S/L}$ gives rise to the  
inclusive  $K^0$ production cross sections 
depicted  in Fig.~\ref{fig:cros_comp}  as function of the
mass of the target nucleus. The systematic
errors (boxes in Fig.~\ref{fig:cros_comp}) are
estimated to be less than 30\%. As determined by extensive 
GEANT-based Monte-Carlo simulations, about 15\% of uncertainty is
related to the evaluation of the reconstruction efficiency and its
dependence on the reconstruction of the primary interaction point,
 whereas about 10\% are due to the chosen selection strategy of
($\pi^-$,$\pi^+$) pairs. An additional error of about 5\% is
attributed to the extrapolation to the full momentum-space; it was
estimated from a comparison to the transport model calculations
described below, which demonstrated that 85\% of the production
cross section can be reconstructed in the geometrical acceptance
of the experiment.

The dependence of the $K^0$ inclusive production cross section on
the mass of the target nucleus, $A$, is fitted with a power law
function: \mbox{$\sigma (\pi^-+A\rightarrow
K^0+X)=\sigma_{\text{eff}}\cdot A^{\text{b}}$}. The result of the
fit yields $\sigma_{\text{eff}}=0.87 \pm$0.13~mb and
${\text{b}}=0.67\pm$0.03, with a statistical error of 
$\chi^2/{\text{ndf}}=0.9/3$. This $A$-dependence suggests that,
in the studied reactions, kaons are
produced at the surface of the nucleus.

{\color{black} Using the observed A-dependence, we compare 
in Fig.~\ref{fig:cros_comp} the measured inclusive cross-sections to 
reference calculations based on the assumption that the production of 
kaons takes place only on the surface of target nuclei with the 
known elementary production cross-sections in vacuum.
The reference values are obtained by summing the  
$\pi^-+p\rightarrow K^0~\Sigma^0$, $\pi^-+p\rightarrow
K^0~\Lambda$ and $\pi^-+n\rightarrow \Sigma^-~K^0$ cross sections weighted with the 
relative neutron and proton contents of the target nuclei and multiplying the results 
by $A^{2/3}$, which represents the effective number of nucleons on the
nucleus surface. 
The results of these calculations are depicted by the 
hatched area in Fig.~\ref{fig:cros_comp}, which includes experimental uncertainties
estimated to be about 20\%. 
The simple scaling ansatz underpredicts the measured cross sections 
by about a factor of 2, which indicates that some essential part of the production 
process is missing. Trivial explanations, like an enhancement of the kaon production 
due to the Fermi motion of the nucleons in the nuclei ~\cite{koptev88}, are ruled out due 
to the weak  dependence of the 
elementary production cross section on incident momentum (see \cite{Tsushima00}).}
{\color{black} 
Applying the  same
scaling procedure as for the vacuum cross-sections to the results of the 
Quark-Meson  Coupling model, which predicts an enhancement of the inclusive kaon production  
at baryon density of   $\rho =\rho_0$~\cite{Tsushima00}, yields the results
depicted by the dashed-dotted line in Fig.~\ref{fig:cros_comp}. 
This type of modification has clearly the potential  
to describe the  magnitude of the observed inclusive cross section.
However, it should 
be noted that the densities for which the calculation is available clearly exceeds 
the density probed by the reactions on surface of nuclei induced by the pion beam -- 
according to the results of the Hadron-String-Dynamics (HSD) transport-model~\cite{Cassing99}, 
kaons are produced on average at densities 
$\rho/\rho_0  > 0.5$, ranging from $0.6$ to  $0.7$ for the 
C\,- and Pb\,-\,target, respectively.
In contrast to the QMC model predictions, available only for
infinite nuclear matter, the results of the HSD calculations, 
can be compared directly to the measured
inclusive production cross-sections (dashed line in  Fig.~\ref{fig:cros_comp}). 
Here, no sensitivity to modifications of kaons can be observed, 
i.e. the calculation without the KN\,-\,potential predicts 
a total production cross-section that is larger by only 3\% than the one 
shown in Fig.~\ref{fig:cros_comp}, which is due to nonresonant 
$K^0$ production-mechanism in the HSD model.}

\begin{figure}[!th]
\vspace{-0.3cm}\includegraphics[width=7.6cm]{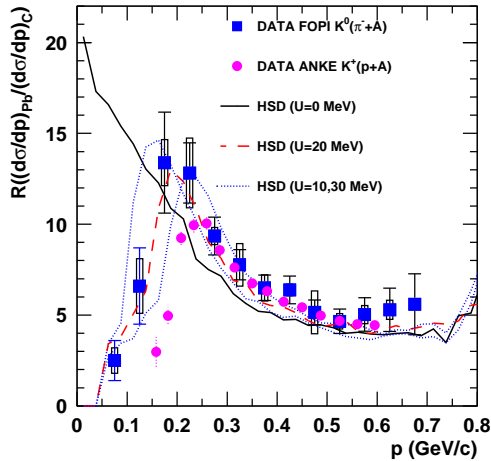}
\caption{\label{fig:rat_comp} The ratio of $K^0$($K^+$) yields
  produced by pions (protons) on heavy and light targets plotted as a
  function of the momentum, $p$, in the lab. system.
  The full squares depict the yield\,-\,ratio of $K^0_S$ produced on
  Pb and C targets in this experiment.
  A similar ratio of $K^+$ yields measured in proton-induced reactions on Au
  and C targets is represented by full circles~\cite{Nekipelov02}. 
  The results of the HSD model 
   with different strength of the KN\,-\,potential are depicted by solid (black,  dashed (red)
   and blue (dotted) lines.}
\end{figure}

{\color{black} 
While, within the present experimental accuracy, the influence of the medium on 
the inclusive  $K^0$ meson production is not 
identifiable when comparing the results to predictions of the HSD model, 
information about the mean-field KN\,-\,potential} can be gained by
comparing the phase-space distributions of kaons produced on heavy
and light targets~\cite{Nekipelov02,Debowski96}.
In particular kaons of low momenta, which spend considerable time 
inside the nuclear matter, are expected to probe the potential efficiently. 
The yields of $K^0_S$ produced on the Pb and the C targets are plotted in
Fig.~\ref{fig:rat_comp} and compared in terms of their ratio as a function 
of the kaon momenta.
The data are corrected for the detection efficiencies, although, 
because of the chosen representation, these correction factors cancel 
out to a large extent. At low momenta ($p<$ 170 MeV/c) the production of
$K^0_S$ on the Pb target is clearly suppressed with respect
to the production on the C target. This observation can be
explained by a repulsive KN\,-\,potential in the nuclear medium,
which accelerates kaons before they escape the nucleus. 
The effect of the acceleration is increasing with the size of the
nucleus because, in case of the Pb target, 
{\color{black} kaons stem from the medium which has a higher density on average.}
In order to estimate the strength of the
potential, the measurements are compared to the results of the HSD transport-model
calculations~\cite{Cassing99}, 
{\color{black} filtered through the geometrical acceptance of the experiment.
The version of the model without the $K^0N$ potential, shown by 
the solid line in Fig.~\ref{fig:rat_comp}, misses the trend in 
the data at low momenta completely. Contrary, the version of the model 
that includes a 20~MeV $K^0N$ potential at $\rho_0$ (with a linear dependence 
of the potential on the nuclear density), depicted by the dashed  
line in Fig.~\ref{fig:rat_comp}, reproduces qualitatively the observed functional dependence 
of the $K^0_S$ yield\,-\,ratio over the full momentum range.}

A similar analysis has been performed by the ANKE collaboration in
the case of $K^+$ meson production by protons of 2.3~GeV energy on
Au and C targets~\cite{Nekipelov02}. At momenta larger than
250~MeV/c, the ratios of $K^+$ mesons yields (full circles in
Fig.~\ref{fig:rat_comp}) agree well with the results of the
present work. At lower momenta, the ratios measured in both
experiments exhibit a similar suppression of kaon production on
heavy targets, but with a different extension of the depletion
region. As in case of the present work, the results of the 
$K^+$ production experiment were reproduced by the transport model calculations
~\cite{Rudy02_Rudy05} with a $K^+$N potential of the order of
20~MeV at $\rho_0$. However, in contrast to positively charged
kaons, the propagation of neutral kaons is not affected by the
(additional) repulsive Coulomb interaction, which in the case of
Au nuclei is as large as 15 MeV. 
Therefore, the strength
of the repulsive KN\,-\,interaction can be extracted generally more
directly from the results of our measurement 
than from previous $K^+$ production experiments. 
{\color{black} The accuracy for the determination of the effective $K^0N$ potential, 
20$\pm$ 5 MeV, was inferred from the comparison of the data to HSD-model 
predictions with different strength of the potential (cf. dotted 
lines in Fig.~\ref{fig:rat_comp}), by means of the $\chi^2$ analysis. 
Experimentally,
this precision is presently limited only by statistics and can be improved by an
intensity upgrade of the SIS18 accelerator.}
A high statistics
experiment would also shed light on the isospin dependence of the
KN potential, a topic that is receiving a high degree of attention
recently \cite{Lopez07}.

In summary, cross-sections of inclusive $K^0$ meson production in
$\pi^-+A\rightarrow K^0~+X$ reactions were measured.
{\color{black} Comparison 
of the results to predictions of the QMC model demonstrates the sensitivity 
of the measurements to the possible changes of   
reaction amplitudes of the underlying
elementary processes at nonzero baryon density. }
The momentum distributions of $K^0_S$ mesons produced on
heavy (Pb) and light (C) nuclei have been compared by means of the 
yield\,-\,ratio and are reproduced by HSD transport model
calculations qualitatively. At low momenta a suppression of $K^0_S$
production on heavy nuclei is observed with respect to the
production on the light target. 
{\color{black} The results of the HSD transport
model calculations suggest that, at $\rho=\rho_0$, a 
repulsive KN\,-\,potential of 20$\pm$ 5 MeV is present due
to $K^0$ interactions with the surrounding nuclear medium.}

We are grateful to the accelerator crew at the GSI facility
for providing the $\pi^-$ beam. This work
was supported by the German BMBF under Contract No.
06HD190I, by the Polish Ministry of Science and
Higher Education under Grant No.\ DFG/34/2007,
by the Korea Science and Engineering Foundation
(KOSEF) under Grant No. F01-2006-000-10035-0, by the
mutual agreement between GSI and IN2P3/CEA, by the Hungarian
OTKA under Grant No.\ 47168, within the Framework
of the WTZ program (Project RUS 02/021), by DAAD (PPP
D/03/44611), and by DFG (Projekt 446-KOR-113/76/04).We
have also received support by the European Commission under
the 6th Framework Program under the Integrated Infrastructure
on: Strongly Interacting Matter (Hadron Physics), Contract
No. RII3-CT-2004-506078.

\end{document}